\documentclass[12pt]{aastex631}
\usepackage{graphicx} 

\usepackage{soul}
\usepackage{float}
\usepackage{parskip}

\begin{document}

\title{Long-slit Spectroscopic Data Reduction For 90-inch Bok Using The Image Reduction Analysis Facility} 
\author[0000-0002-0786-7307]{Sara E. Holeman}
\author[0000-0002-0786-7307]{Sanchayeeta Borthakur}
\affiliation{ School of Earth \& Space Exploration \\
Arizona State University \\
781 Terrace Mall, Tempe, AZ 85287, USA.}
\cleardoublepage 



\setcounter{page}{1} 

\section{\textbf{Introduction}}\label{sec:intro} 

This manual is specific to the procedure of reducing galaxy long-slit spectroscopic data from the Boller \& Chivens Spectrograph aboard the 90” Bok telescope at Kitt Peak National Observatory. The Image Reduction Analysis Facility (IRAF) is utilized to complete the steps of image reduction. This manual will discuss methods of removing instrumental signatures, correcting for radiation events from environmental conditions, and the data frames for wavelength calibration for the particular data presented. We refer the reader to \textit{A User's Guide to CCD Reductions with IRAF} [1] for more general and in-depth information.

In Section~\ref{sec:spectr_format}, 
the spectra data format and calibration frames will be introduced. Section~\ref{sec:iraf_red}, will present IRAF tasks with their appropriate parameters following the steps of the initial reduction. Lastly, Section~\ref{sec:wav_cal} will highlight the importance of wavelength calibration with the tasks that create a two-dimensional function to project the scientific data on a linear wavelength scale. 
\\

\section{\textbf{Spectra Format \& Calibration Frames}}\label{sec:spectr_format}

Upon collecting data, several artificial components are superimposed on the CCD due to instrumental factors such as the electronic bias level, the multiplicative gain and illumination variation, and any non-linear radiation events, particularly cosmic rays. Therefore, removing these fallacious data components is required before any scientific data analysis.

\subsection{Two-dimensional galaxy spectra} \label{subsec:2-d_gal_spec}

\begin{figure}[h]
  
         \includegraphics[scale=.35]{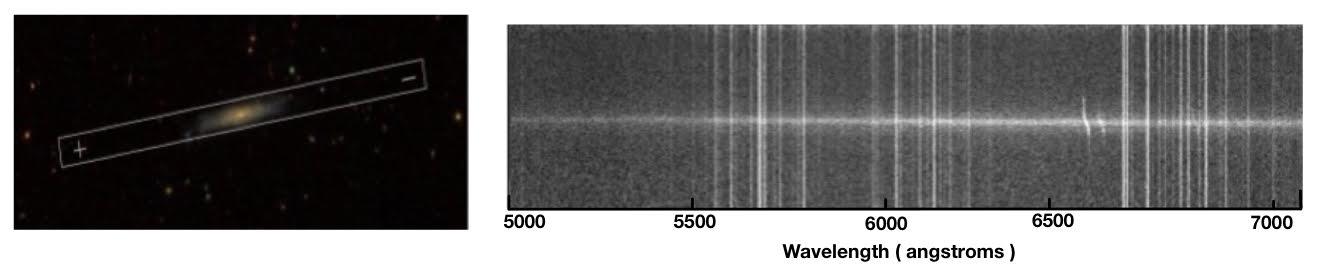} 
         \centering
         \caption{Left: Panel showing the position of the slit with respect to the target galaxy. Right: The two-dimensional data showing the galaxy continuum. The H$\alpha$ emission line can be clearly identified thus indicating the rotation of the galaxy by its red and blue shifted wavelengths. The vertical lines along the frame are emission skylines.}     
    \label{fig:slit2draw}
\end{figure}

Analyzing a two-dimensional spectrum, you will see the flux as a function of wavelength. Reference {{ Figure~\ref{fig:slit2draw}}} for the slit position relative to the major axis of the galaxy and reduced two-dimensional spectra of a low-redshift galaxy, showing prominent H$\alpha$ emission lines extending out to the galaxy disk.

\subsection{Overscan Region \& Bias Frames} \label{subsec:bias_frames}

While obtaining the data, an additional voltage is added to the output signal, that comes from the CCD's sensor while reading out the signals from light exposure. Therefore, the additional counts need to be subtracted from all the frames as the initial step of the image reduction. This step is referred to as ``bias subtraction'' and leads to identifying the overscan region where no incident light falls onto it during exposures. The overscan region, which exists at the tail end of the CCD (around 30 pixels in length and the full width of the CCD), allows only the biased voltage to be defined. As this value varies slightly along the columns of the CCD, a first-degree polynomial fit is easy to obtain or simply averaging the data over the columns of the overscan region is sufficient as well. The overscan region will be trimmed off after subtracting the fit from your images. 

\vspace{-.5cm}

It is typical to expose the CCD to a flash of uniformed light (pre-flash) to ensure no charge loss during signal transfer. This superimposes yet another number of counts on the data that the bias subtraction may not account for. This can be corrected by combining a collection of bias frames. These frames are ``zero'' second integration exposures, and by median combining to a single frame, the small pre-flash counts can be subtracted from the other reduction frames. 
 
\subsection{Dark Current} \label{subsec:dark_cur}

Dark current is a temperature-dependent source of additional noise accumulating over long exposures. In the case of the data presented here, the dark current is not significant and, therefore, not corrected. However, it is possible to check if correction is needed. Because the overscan and bias levels would have already been removed from all calibration frames, if any counts are visible on the dark frames, then that will be a significant source (i.e. real) of dark current.

\subsection{ Flat field correction \& Fixing non-linear pixel sensitivities} \label{subsec:flat_fixpix}

Flat field data is fundamental to help define light sensitivities along the CCD, it is an image of uniformly distributed light. Dividing the raw data by a normalized flat field, (done in an IRAF), corrects for uneven illumination, dust, and vignetting in an image. The preservation of counts is important for spectra analysis. Thus, two additional requirements must be met before normalization. Firstly, it is useful to interpolate over any bad pixels on the CCD, which ultimately appear in the images. Interpolation can be done by creating a bad pixel mask that will be used during the final correction of the reduction process. Second, possible wavy distortions may appear on the images showing a large-scale variation in light that is wavelength-dependent caused by several potential factors like the temperature or color imbalances of the flat-field lamp. Two tasks will be used for both processes and introduced in the following section. 
\\

\section{\textbf{Image reduction with IRAF }} \label{sec:iraf_red}

IRAF is a command language program with packages that perform tasks depending on the specific parameters given. This section will cover image reduction package tasks and instructions on the parameters that are used to correct raw data. To edit the parameters of a task enter \textbf{epar} in the command line followed by the task name you'd like to edit (e.g., \textbf{epar zerocombine}). Once the parameters are appropriate to the step at hand, save them by \textbf{control-Z} on your keyboard.

\subsection{Characterizing the data} \label{subsec:characterizing_data}

\begin{figure}[H]
    \centering
    \includegraphics[scale =.6]{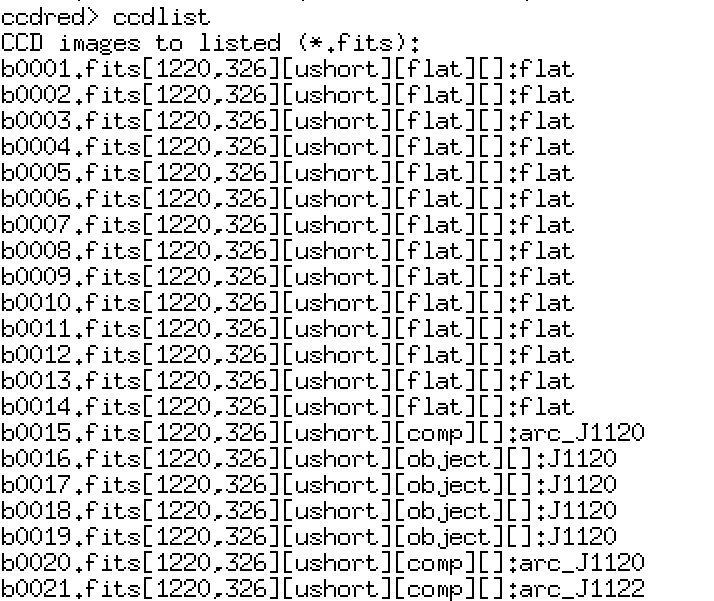}
    \vspace{0.25cm}
    \caption{\textbf{ccdlist} will output a list shown above presenting the header information for each frame in the dataset regarding its dimensions with its data and object type.}  
    \label{fig:ccdlist.png}
\end{figure}

To familiarize yourself with the data, use the task \textbf{ccdlist} to analyze the information of individual frames present in the folder you are working in. The output can be seen in Figure~\ref{fig:ccdlist.png}, where each frame has header information associated with the object type defined by the observer at the telescope. A translation table regarding the header information must be set up so that IRAF can reference the appropriate data when running called tasks. The \textbf{setinstrument} task will do so and is located in the \textbf{ccdred} package. For spectral reduction, the input parameter needs to be set \textbf{specphot}. Upon running this task, you will enter another parameter editor for the \textbf{ccdred} package. Note that the primary defaults of the package will work just fine but ensure that the pixel type parameter is set to \textbf{real}.

\subsection{Locating the overscan region} \label{subsec:loc_over}

\begin{figure}[b]
    \centering
    \includegraphics[scale=0.25]{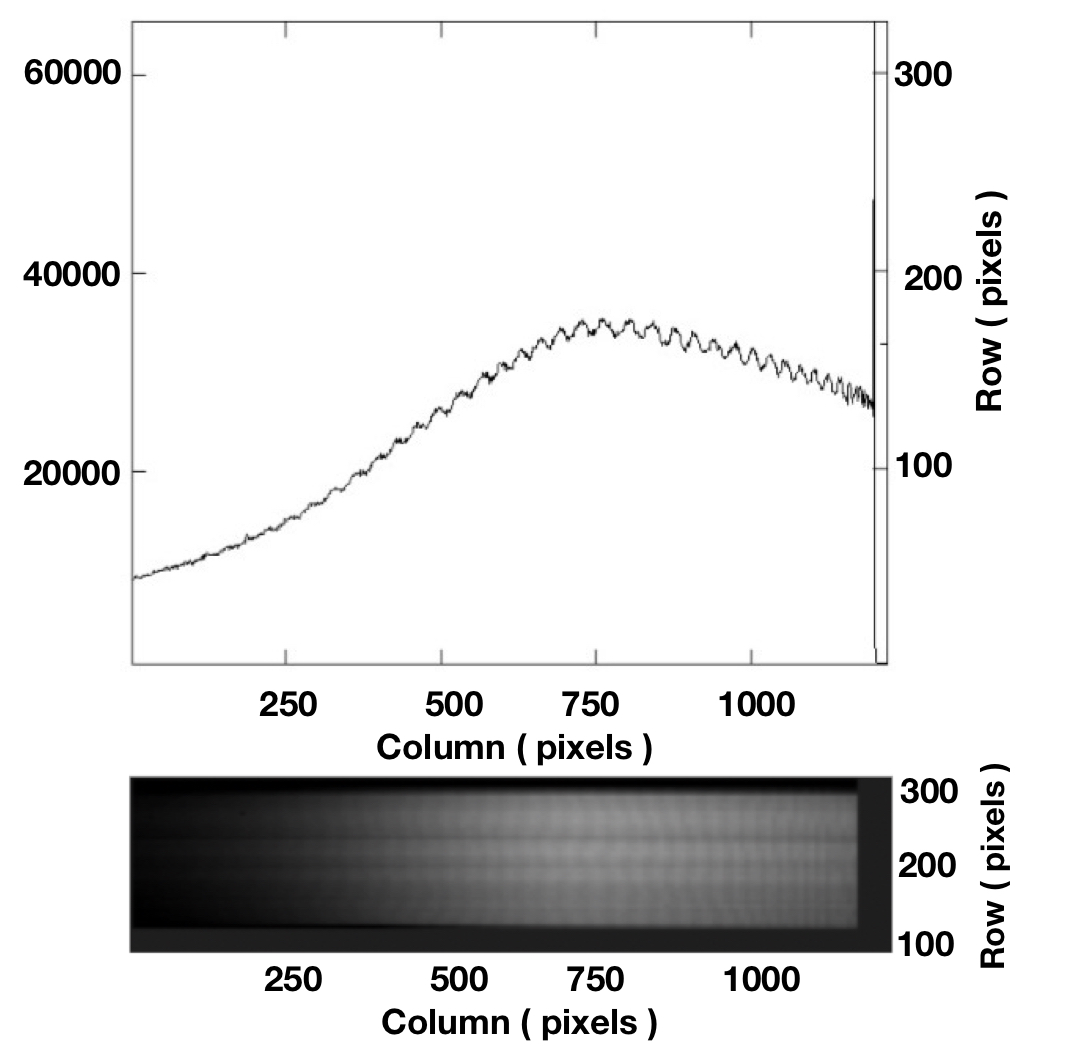}
    \vspace{-0.2cm}
    \caption{Top: Flat field plot showing the light variation across the CCDs pixel columns.  Bottom: For visual inspection, the overscan region is the dark portion of the CCD where there's no penetrating light (see far right, top, and bottom of the frame). }  
    \label{fig:overscandraw}
\end{figure}

Examining the flat field data can be useful in identifying the location of the overscan region. There are two ways of defining this region. First, by visual inspection and locating the specific column where the cutoff is along the CCD and, secondly, by examining the plot of the flat field with the IRAF \textbf{implot} task. Figure~\ref{fig:overscandraw} presents the two forms we can visualize the data.

\clearpage

\subsection{Combining Bias Frames with \textbf{zerocombine}} \label{subsec:zerocomb}

\begin{figure}[h]
    \centering
    \includegraphics[scale=0.38]{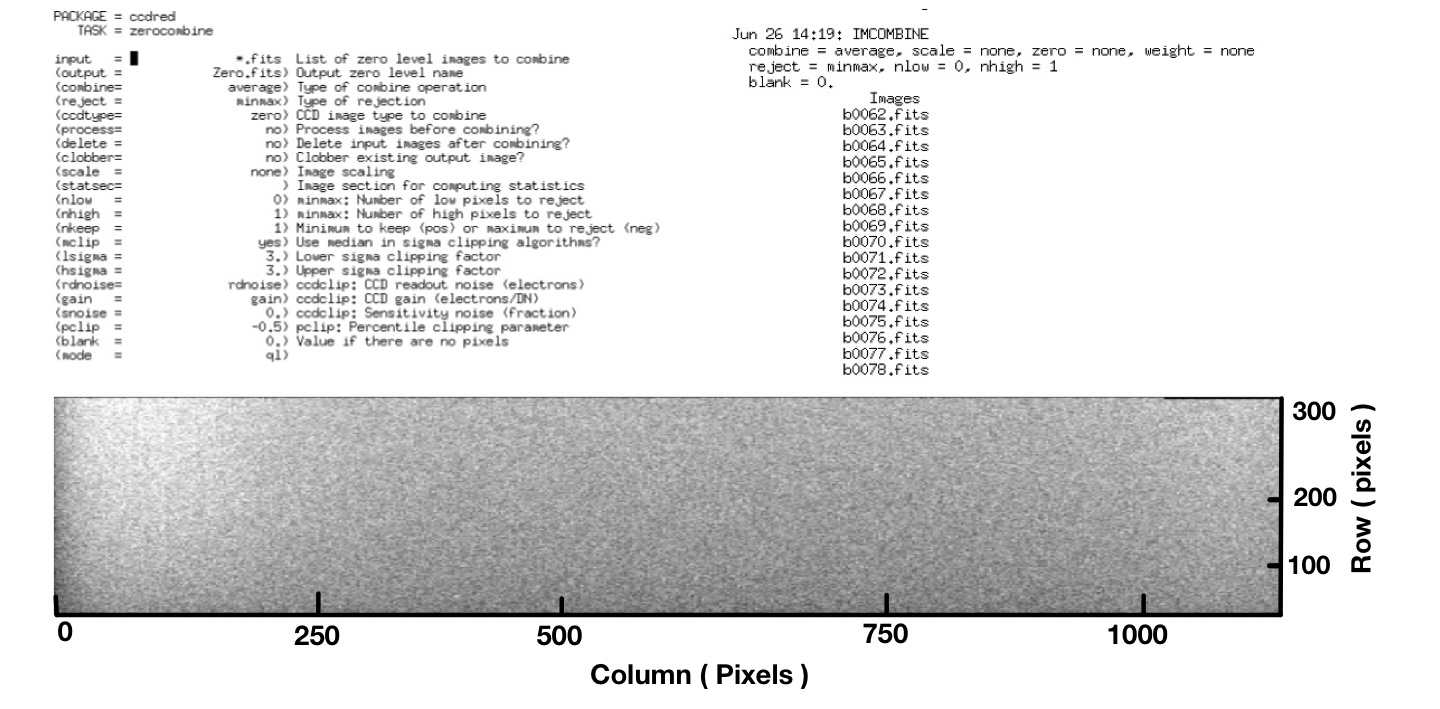}
    \vspace{-0.4cm}
    \caption{ Top Left: Initializing the \textbf{zerocombine} task with parameters to combine all bias frames by a \textbf{minmax} average thus rejecting the minimum and maximum value in each pixel and producing a ``master bias'' fits file titled Zero.fits. Top Right: The expected output IRAF displays when the combination of selected frames. Bottom: The final combined ``master bias'' titled ``Zero.fits''. }  
    \label{fig:zerobias}
\end{figure}

To ensure no charge loss during signal transfer, an additional number of counts are added to the data by pre-flashing the CCD before exposures. Combining data frames is a technique that is frequently used in this manual. The combined output image is sometimes called a ``master'' of whatever frames you combined, for example, a ``master flat'' or ``master bias''. The task for combining the bias data is \textbf{zerocombine} with the altered input parameters and expected output shown in Figure~\ref{fig:zerobias}. Notice that we will use an average combination operation and identify the object type as ``zero'' which tells IRAF to collect the bias frames using their header information. Running the task will create a single average combined biased frame titled ``Zero.fits'' as defined as the output parameter. Reference the bottom frame of Figure~\ref{fig:zerobias} to see the combined ``master bias''. Creating a master frame will reduce the random noise from the CCD sensor, it will smooth out the pattern of noise providing a better depiction of the bias level that will be subtracted.

\vspace{-.5in}

\subsection{ Initializing \textbf{ccdproc} with bias correction and trimming the data } \label{subsec:init_ccdproc}

The task \textbf{ccdproc} is helpful in image reduction and will be run twice. The first run will perform bias subtraction, preflash correction, and trim a defined section of the CCD with the data wanted for analysis. Note that utilizing as much of the CCD is recommended for long-slit data. Therefore, in the \textbf{trimsec} parameter, the columns and rows defined as \textit{[x1:x2,y1:y2]} will be the portion of the CCD for analysis; any region not defined will be cut off. A benefit of IRAF, is that it provides the option to alter the functions we are trying to fit to our data.  When \textbf{ccdproc} is run, IRAF will ask in the command line if you want to fit the data interactively. By answering ``yes'', you will enter the interactive curve fitting mode where you can change the fitting parameters. Analyzing a low rms and editing the order of the fit is dependent on preference, however, the default settings should be sufficient. Enter ''q`` to save and exit the fitting mode. If happy with the fit, answer ``NO'' (in all caps) to the same question in the command line. Once completed, the section defined will be trimmed off and the bias level removed.

\begin{figure}[H]
    \centering
    \includegraphics[scale=.35]{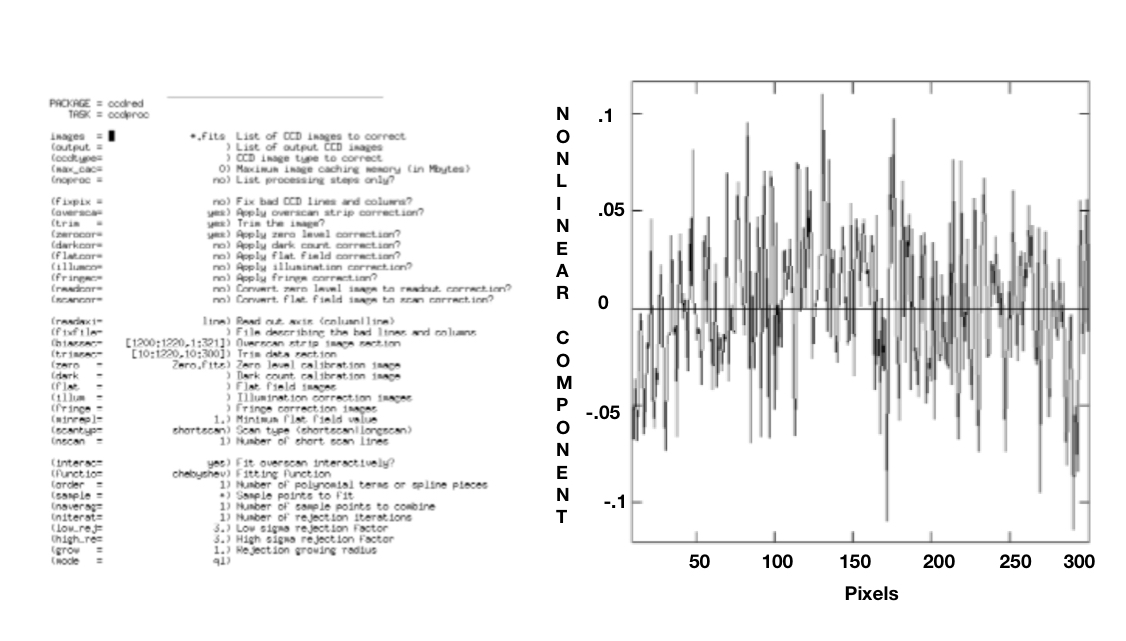}
    \vspace{-0.4cm}
    \caption{ Left: Initializing the \textbf{ccdproc} with default parameters will create a fit to the bias level of first or second order. Right: An example of the bias level across ``Zero.fits''.}  
    \label{fig:ccdprocdraw}
\end{figure}

\vspace{6pt}

\subsection{ Initializing the \textbf{fixpix} procedure } \label{subsec:fix}

\begin{figure}
    \centering
    \includegraphics[scale=0.8]{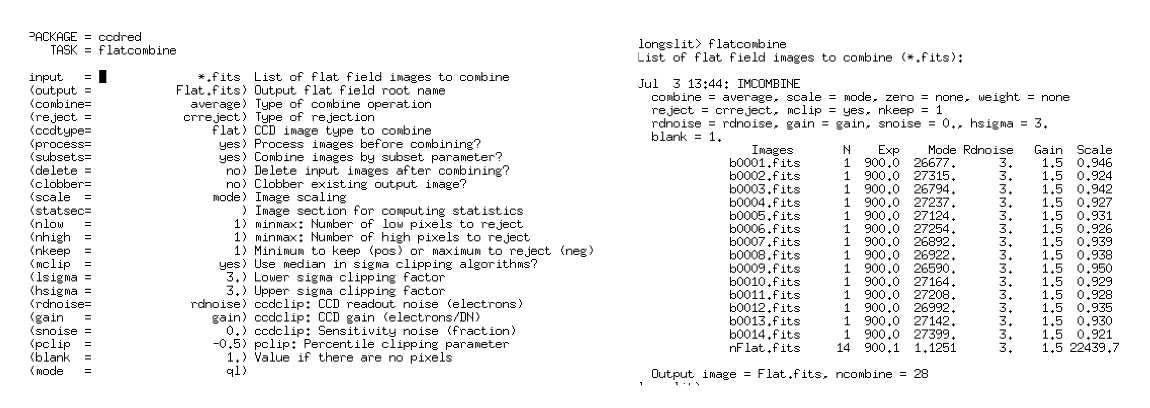}
    \vspace{-0.4cm}
    \caption{ Similar to the ``Zero.fits'' master frame, a ``master flat'' is created. Left: rejecting any high pixel values caused by the various cosmic rays that fall on the CCD during exposures. Right: Expected output of the combined flat field frames, creating the master frame title ``Flat.fits''.}
    \label{fig:initializingfixpix}
\end{figure}

Corrupted pixels must be interpolated over with the use of a mask created from the combination of multiple flat field exposures. It is easiest to identify these regions from the flat data. \textbf{fixpix} will be the final task that will do interpolate over the bad pixels but needs the combined flat field data.  The task for combining flat field data is \textbf{flatcombine} with the altered input parameters and expected output shown in Figure~\ref{fig:initializingfixpix}. Changing the \textbf{reject} option to \textbf{crreject} will correct for potential radiation events like cosmic rays. This new ``master flat'' will be used to create the bad pixel mask by entering it's output name ``Flat'' as the input of \textbf{ccdmask} task. The default parameters for \textbf{ccdmask} do not need to be changed. Once the mask is created, it should be stored in a separate folder so that it does not get altered by the upcoming tasks. The mask will be used as the final step of the reduction process. 

\subsection{ Flat correction  with \textbf{response}} \label{subsec:flatcor}
To perform the flat field correction, a ``master flat'' must be created again. Flat field correction for spectra data has an additional step when normalizing the ``master flat'' that imaging does not need. Therefore, a function will be fitted to the data along the wavelength axis. To access the task needed to perform the removal and normalization, call the primary package \textbf{noao} in the command line, then to access the additional packages needed, enter \textbf{twodspec}, then subsequently after \textbf{long-slit}. The \textbf{response} task and default parameters will produce an output image that will be the ratio of the normalized flat frame to the fit created. The function of the fit can be changed to a cubic spline of order 6 by entering \textbf{:funct spline3} and \textbf{:order 6}. Now the \textbf{ccdproc} task can be run once again ( taking into account that it was already run once for the overscan and bias corrections) to perform the flat-field division by entering the output image created with \textbf{response} as the \textbf{flatcor} input parameter.

\vspace{8pt}


\subsection{ Final Correction  with \textbf{fixpix}} \label{subsec:tables}

The final correction in the image reduction will be to interpolate over the bad pixels on the CCD. To do so, access the bad pixel mask initially created and stored in a separate folder. Enter it as the input for the \textbf{fixpix} task along with the set default parameters. This task will take a few seconds to run, and once completed, a new command line will start, thus concluding the image reduction procedure. 
\\

\section{\textbf{Introduction to Wavelength Calibration}}\label{sec:wav_cal}

Wavelength calibration is a process that places the reduced two-dimensional spectra on a linear wavelength scale. A fit is created from a comparison spectrum with laboratory wavelengths and is applied to the reduced two-dimensional spectra. The process will be discussed with the IRAF tasks from the \textbf{twod.longslit} package. 
\vspace{6pt}

\subsection{ Arc Frames } \label{subsec:arcdraw}

\begin{figure}
    \centering
    \includegraphics[scale=.35]{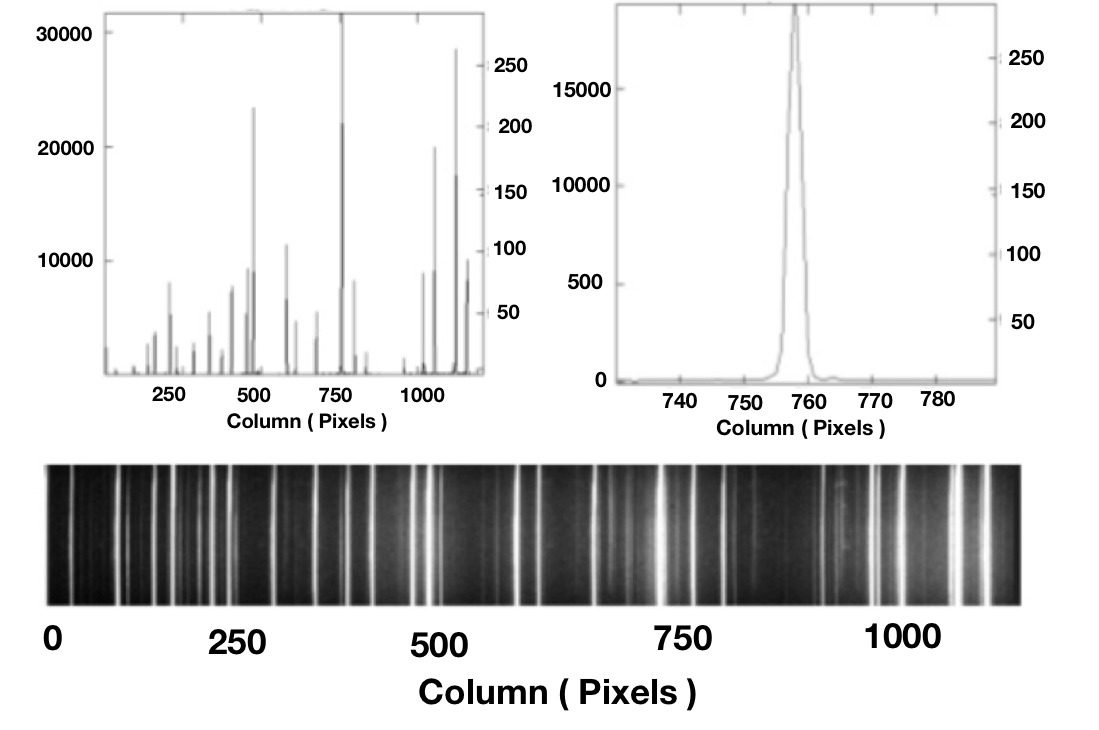}
    \vspace{-0.2cm}
    \caption{ Top Left: The emission lineplot that displays the comparison spectra of the HeNeAr lamp used for wavelength calibration. Top Right: Narrowing in on a single line lets the FWHM be defined for the use of an input parameter for future tasks performed for the calibration process. Bottom: The two-dimensional image of the data from the discharge lamp. }  
    \label{fig:arcdraw} 
\end{figure}

An arc or comparison frame is produced by the illumination of a gas-discharge lamp where the light penetrates ionized gas of various elements, thus creating well-characterized emission line spectra. These frames allow for identifying known wavelength measurements as reference for the two-dimensional data. For the long-slit data presented in the left panel of  Figure~\ref{fig:arcdraw}, a Helium, Neon, and Argon lamp (HeNeAr) was used at the Bok telescope to produce the comparison spectra. By narrowing in on a single line like on the right of Figure~\ref{fig:arcdraw}, one can help define the line full-width half max (FWHM) for the upcoming wavelength calibration procedure. Analogous to the two-dimensional data shown in  Figure~\ref{fig:slit2draw}, the illumination pattern that the HeNeAr lamp produces is shown at the bottom panel. 

\subsection{ Identifying emission lines } \label{subsec:emission_line}
\begin{figure}
    \centering
    
    \includegraphics[scale=0.45]{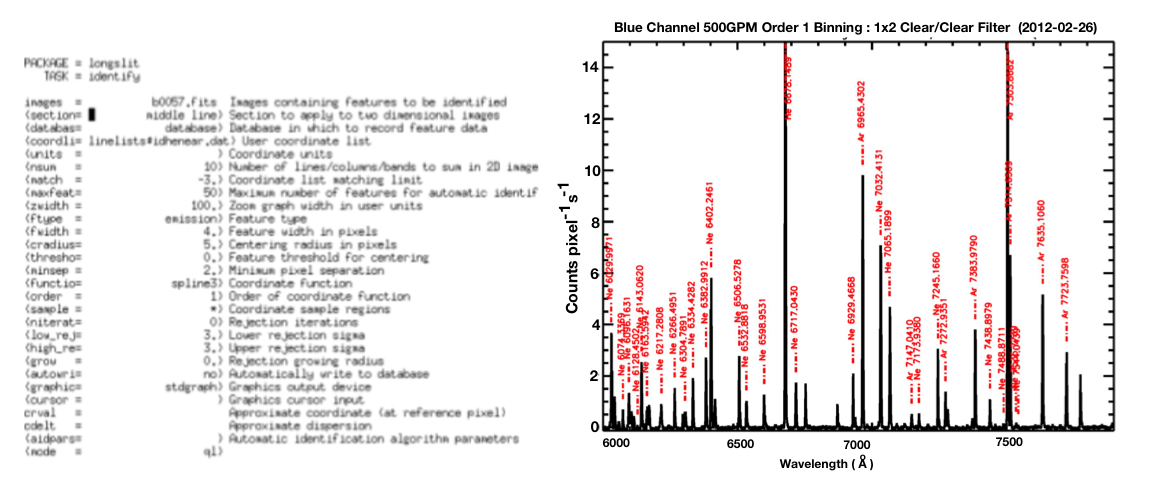}
    \vspace{-0.20cm}
    \caption{ The identification of emission lines. Left: Presents default parameters of \textbf{identify} used with either a single arc frame or a combined arc, depending on the data you have collected. Left: The comparison plot given for convenience to find the exact value of the emission lines you want to define for the preliminary fit of a dispersion solution. }  
    \label{fig:emissiondraw} 
\end{figure}

The first step is identifying the known wavelengths from the comparison spectrum to create a fit of the lines. This will be done using the \textbf{identify} task. Shown in {{Figure~\ref{fig:emissiondraw}} to the left is the \textbf{identify} parameters where the input will be the comparison frame as well as what lamp was used in the \textbf{coordlist} parameter and also the FWHM pixel value. By entering the specific lamp, IRAF can access its reference table, which reduces the number of lines to identify manually to about two or three. To identify lines, reference the HeNeAr reference plot in Figure~\ref{fig:emissiondraw}.

Choose three lines to define by selecting each with the \textbf{m} key and enter the appropriate wavelength (in Angstroms) shown in the top left of Figure~\ref{fig:identify}. IRAF will attempt to identify the element based on the provided wavelength. It is important to verify that this was done correctly. If not, delete the mark you made by hovering the cursor over the line and entering \textbf{d}. Once a few lines are defined, a preliminary fit will be created when the \textbf{f} key is entered. Return to the identification portion of the task by entering \textbf{q}. Creating this initial fit lets IRAF reference and identify all other wavelengths the moment \textbf{l} is selected. Here, it should be clear that the marks are centered on each line. Create another fit for all lines and perform any necessary changes to receive a low residual level. Here a \textbf{:funct spline3} of \textbf{:order 3} will suffice. Exiting out with \textbf{q} and answering ``yes'' to the question ``Write to the database?'' will save the fit to a separate database folder for access later on.


The \textbf{Reidentify} task uses the first dispersion solution from the database as a reference spectrum to ensure that no lines are lost and to account for any misalignments in the lines themselves. Therefore, \textbf{reidentify} will look for the same wavelengths for every ten lines along the spatial axis and create a new fit with any corrections made. The input to the top and its output at the bottom of in Figure~\ref{fig:reidentifydraw} can be referenced once answering ``yes'' yet again to the question ``Write to the database?'' where it saves the redefined fit to the separate database folder.
\begin{figure}[!t]
    \centering
    \includegraphics[scale=0.3]{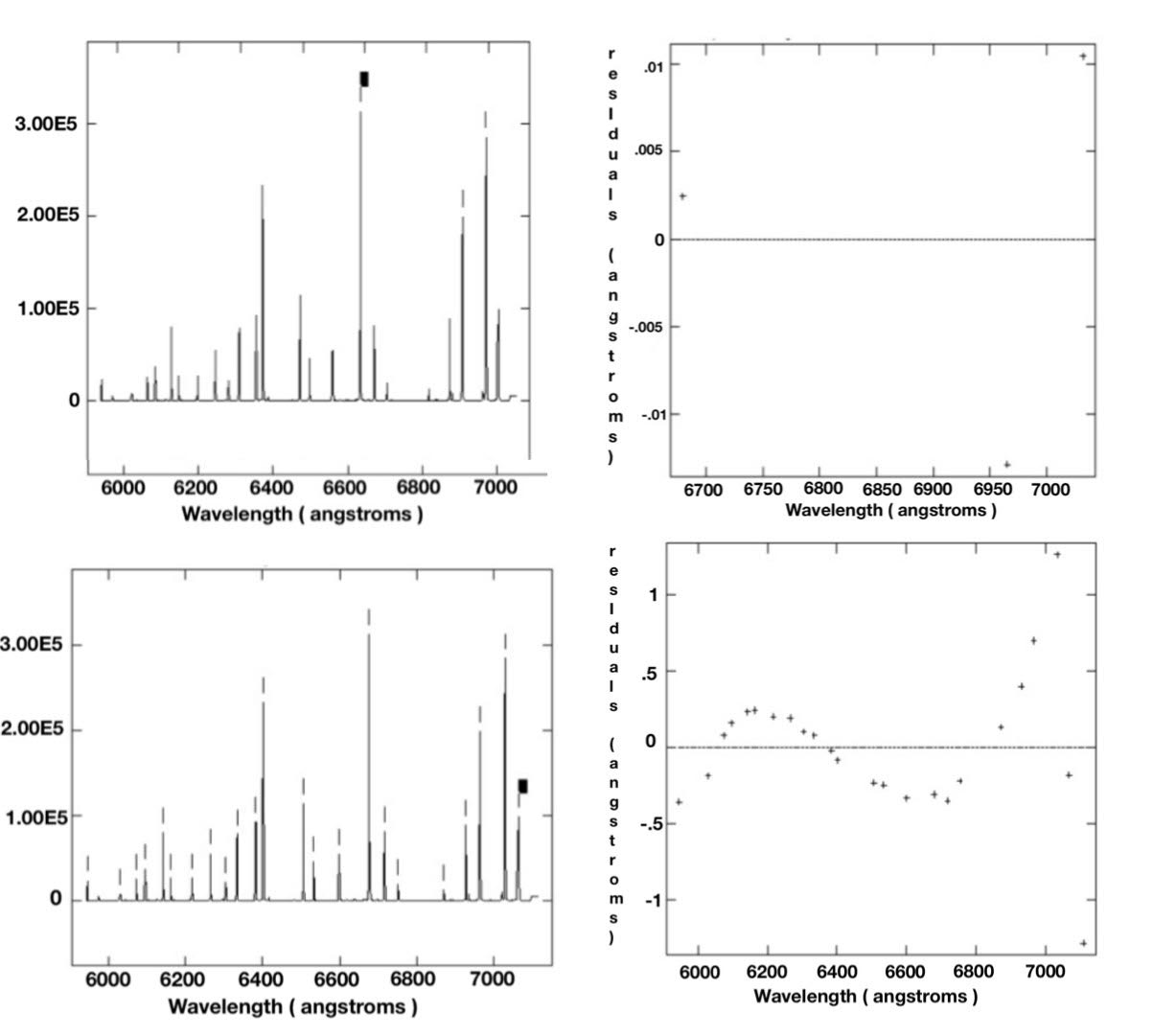}
    \vspace{-.2cm}
    \caption{Top Left: Initially defining a preliminary fit with a few lines chosen. Top Right: With three lines the preliminary fit is created with a standard ordered fit to let IRAF thus us \textbf{identify} determine all of the lines. Bottom right: With all lines defined a better fit is used for the dispersion solution and saved in a database. }  
    \label{fig:identify} 
\end{figure}

\begin{figure}
    \centering
    \includegraphics[scale=0.8]{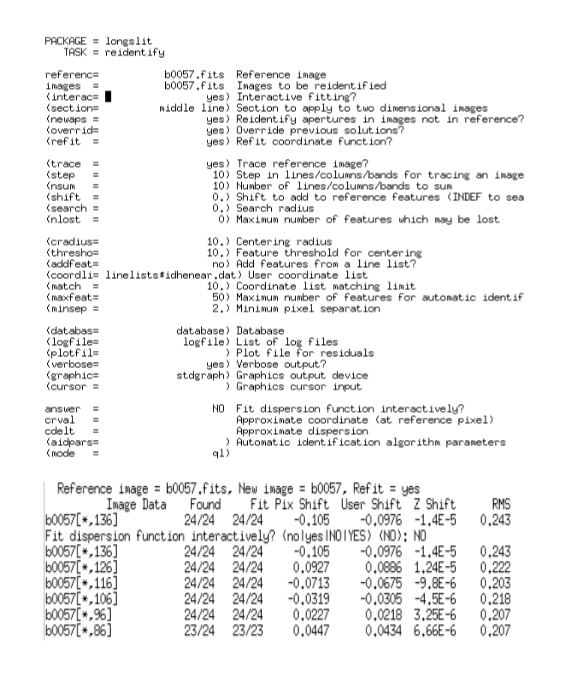}
    \vspace{-0.4cm}
    \caption{ Top: Default parameters of the \textbf{reidentify} task that will collect the preliminary fit created to double-check lines. Bottom: the output information indicating the RMS values for each line fits created. }
    \label{fig:reidentifydraw}
\end{figure}

\vspace{.60cm}
\subsection{ Defining coordinates with \textbf{fitcoords}} \label{subsec:fitcoords}

Next, a two-dimensional fit must be created to provide a full wavelength solution for the image data. Thus, utilizing the \textbf{fitcoords} will allow for inspection of the residual level with respect to the x and y axis separately. A few keyboard sequences of selecting \textbf{x}, \textbf{y}, and \textbf{r} can show each axis respective to its residuals. When selecting a sequence, entering \textbf{r} will replot the data with the chosen sequence. This allows for examining the linear fit for the two-dimensional dispersion solution. Each subfigure presented within Figure~\ref{fig:fitcoordsdraw} will show possible key sequences with the order chosen for the lowest residual values for the corresponding fit. If there are any points along the rows or columns that need to be deleted, do so by typing \textbf{d} followed by \textbf{p} then \textbf{f} to have a corrected fit. Yet again, once happy with the overall fit per axis versus residual, save the fit by exiting \textbf{fitcoords} with the \textbf{q} key. 

\begin{figure}
    \centering
    \includegraphics[scale=0.5]{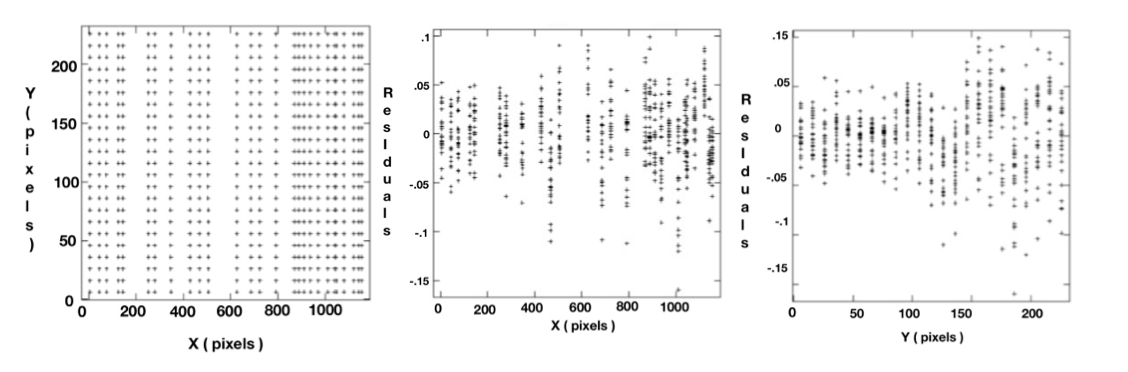}
    \caption{Inspecting each fit along the different axes with various key sequences will make sure that the fit has a low RMS for each sequence respectively following left to right along the three panels, (x,y), (x,r), and (y,r).} 
    \label{fig:fitcoordsdraw}
\end{figure}

\vspace{.1cm}

\subsection{ Coordinate transformation with \textbf{transform}  } \label{subsec:transform}

Lastly, projecting the wavelength solution onto the scientific data will be done with the \textbf{transform} task and setting the input parameter to the scientific frame of interest and its corresponding comparison spectra. You may choose whichever name to define as the output frames. Thus, running \textbf{transform} will collect the two-dimensional wavelength solution created from the database subdirectory and apply it to the input frames chosen. 

\subsection{ Optional redshift correction with \textbf{dopcor}} \label{subsec:dopcor}
An additional task can be performed and may be useful for extragalactic spectra. By knowing the systemic redshift of the target, the \textbf{dopcor} task can redshift correct the spectra so that any specific emission lines can be shifted to their rest-frame wavelengths.

\section{\textbf{Concluding statements}}\label{sec:conclusion}

An example of the final, reduced 2D spectrum is shown in Figure~\ref{fig:spectraspatial}. You may notice that emission skylines are no longer present; a developed Python code removed them. However, the task of background removal and spectra extraction using IRAF can be referenced in \textit{A User's Guide to Reducing Slit Spectra with IRAF} by Phil Massey [2]. Note that skyline removal can be challenging. Thus, it is expected that there will be over-subtraction (darker lines down the location of stronger lines) on the final image. The corrected two-dimensional image can be collapsed into a spatial profile by summing the pixel values along the spatial axis, presented in Figure~\ref{fig:spectraspatial}. Figure~\ref{fig:final} presents the extracted spectra with a fit (seen in red) on the H$\alpha$ emission line, which is present in the two-dimensional scientific data. The spectra presented shows the H$\alpha$ line that's present in an aperture that's placed along the slit of the galaxy disk.

\begin{figure}[H]
    \includegraphics[scale=.45]{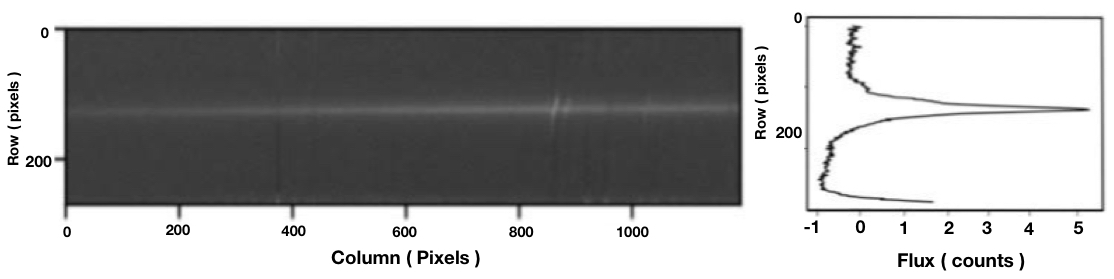} 
         \caption{What to expect for spectra analysis of a, Left: fully reduced and wavelength corrected two-dimensional spectra of a galaxy with H$\alpha$ emission lines. Right: An example of analysis regarding the spatial profile of the galaxy.} 
    \label{fig:spectraspatial}
\end{figure}

\begin{figure}[H]
    \includegraphics[scale=.35]{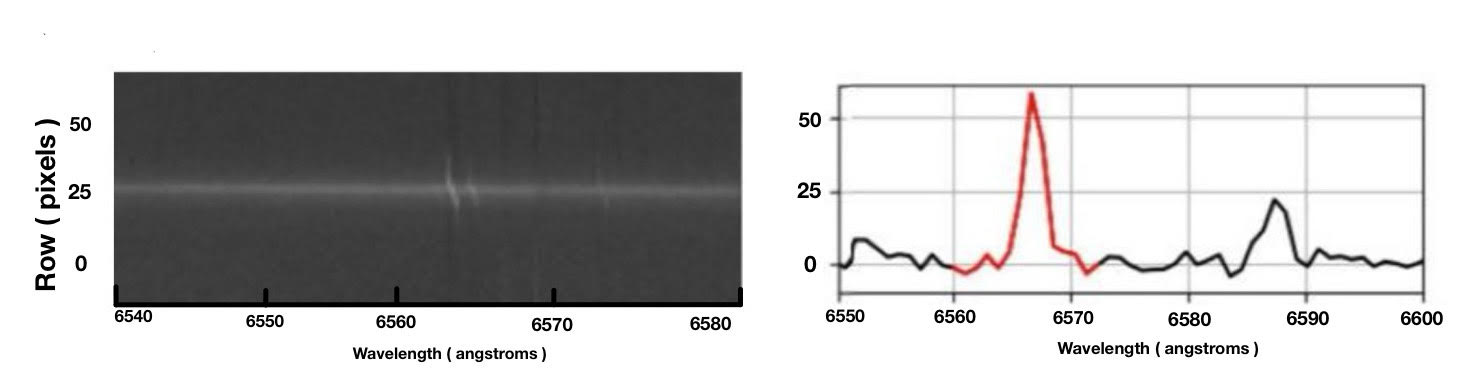} 
         \caption{The narrowed region on the H$\alpha$ region of the two-dimensional scientific data and the extracted spectra. In red is a fit specific to the H$\alpha$ line } 
    \label{fig:final}
\end{figure}

\section{\textbf{\bf Acknowledgments}}

Sara Holeman and Professor Sanchayeeta Borthakur acknowledge the support for this work through the National Science Foundation grants 2108159 and 2408050. Sara Holeman thanks the NSF support through the REU program associated with the grants mentioned-above.

We thank Professor Paul Smith for his input during observations at the Bok telescopes. We also thank the staff at Bok telescope, Steward Observatory for their help and support. 
We also thank Chris Dupuis, Alejandro Olvera, Mansi Padave, Jackie Monkiewicz, and Brad Koplitz for their advice on observations and comments on this manuscript.

\section{\textbf{References}}\label{sec:refs}

[1] Massey, Phillip. (1997). A User’s Guide to CCD Reductions with IRAF. Retrieved 2024, from https://iraf-community.github.io/doc/iraf.pdf. 

[2] Massey, P., Valdes, F., \& Barnes, J. (1992, April 15). A user’s guide to reducing slit spectra with IRAF.         

\end{document}